\documentclass[11pt,twoside]{article}
\usepackage{asp2010}
\usepackage{natbib}

\resetcounters

\begin{document}

\label{authorguide}

\allowtitlefootnote

\title{Simulating the circum-stellar environment of supernova and GRB progenitors by combining stellar evolution models and hydrodynamical code}
\author{C.~Georgy, R.~Walder, and D.~Folini}
\affil{Centre de Recherche Astrophysique de Lyon, CRAL\\
Ecole Normale Sup\'erieure de Lyon, ENS\\
46, All\'ee d'Italie, 69364 Lyon -- France}

\begin{abstract}
The medium around massive stars is strongly shaped by the stellar winds. Those winds depend on various stellar parameters (effective temperature, luminosity, chemical composition, rotation, ...), which are varying as a function of the time. Using the wind properties obtained by classical stellar evolution code allows for the multi-D hydrodynamical simulation of the circum-stellar medium accounting for the time variations of the wind. We present here the preliminary results of the simulation of the medium around a fast rotating star. Comparing the results of such simulations with the observed properties of the circum-stellar medium will allow to better understand the interactions of the stellar winds with the interstellar medium, and could give hints on the past history of the mass loss around massive stars, providing constraints for stellar models.
\end{abstract}

\section{Mass loss by fast rotating massive stars}

It is known from a long time that the luminous flux at the surface of a rotating star is not constant but varies with the colatitude \citep{vonZeipel1924a}. This result was generalised to the case of stars presenting a ``shellular''-like rotation \citep{Maeder1999a}. It follows that the luminous flux is stronger at the poles than in the equatorial regions. According to the so-called CAK-theory \citep{Castor1975a}, the mass loss from hot massive stars is governed by the interaction of the luminous flux with the spectral lines. One can thus expect an anisotropic mass loss around such stars \citep[see e.g.][]{Maeder2002a}.

This anisotropic mass flux around the star modifies the amount of angular momentum removed from the surface. However, the account for the anisotropic winds in stellar evolution simulations produces only minor differences compared to models where this effect is not accounted for \citep{Georgy2011a}. The effect of such winds on stellar evolution is thus negligible. One can however wonder whether the anisotropic stellar winds can modify the circum-stellar medium (CSM), and eventually leave some imprints which could be observable around fast rotating massive stars, or during the supernova (SN) explosion.

In order to explore this direction, we study the case of a very fast rotating $20\,\mathrm{M}_\odot$ star at very low metallicity ($Z=5\cdot 10^{-5}$). As we are interested in the anisotropic winds, we want that the model rotates as quick as possible. The initial angular velocity is set at $\Omega_\mathrm{ini} = 0.7\Omega_\mathrm{crit}$ where $\Omega_\mathrm{crit}$ is the critical angular velocity, defined by the vanishing of the equatorial effective gravity under the effect of the centrifugal force \citep{Maeder2000a}. The low metallicity chosen implies a weak mass loss, ensuring a weak braking of the stellar surface. Moreover, the stellar model is computed accounting for the internal coupling due to magnetic field \citep{Maeder2004a}, also favouring fast rotation. The model is computed with the Geneva stellar evolution code \citep{Eggenberger2008a} from the zero age main sequence up to the end of the central silicon burning.

\begin{figure}
\includegraphics[width=.4391\textwidth]{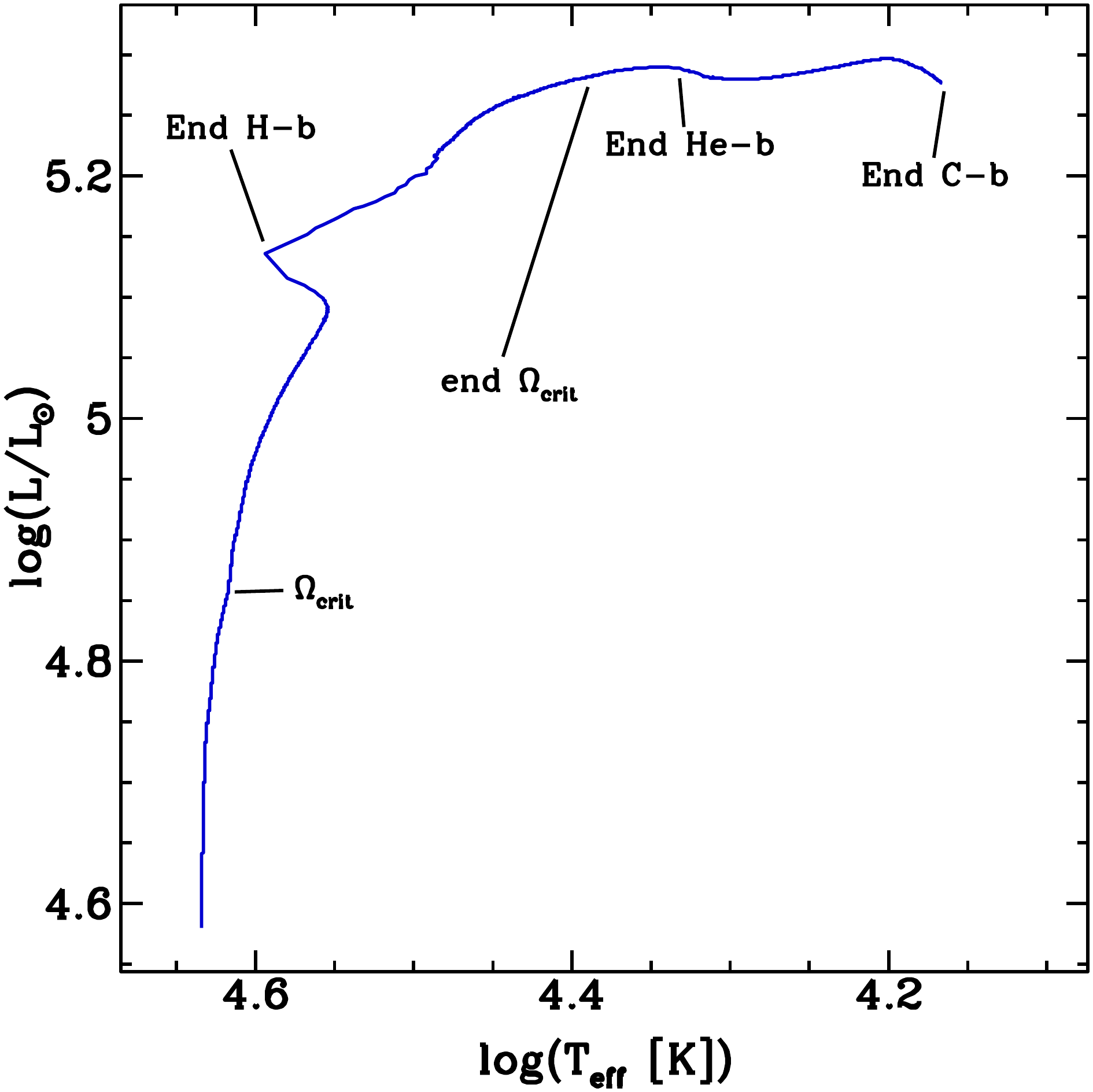}\hfill\includegraphics[width=.5439\textwidth]{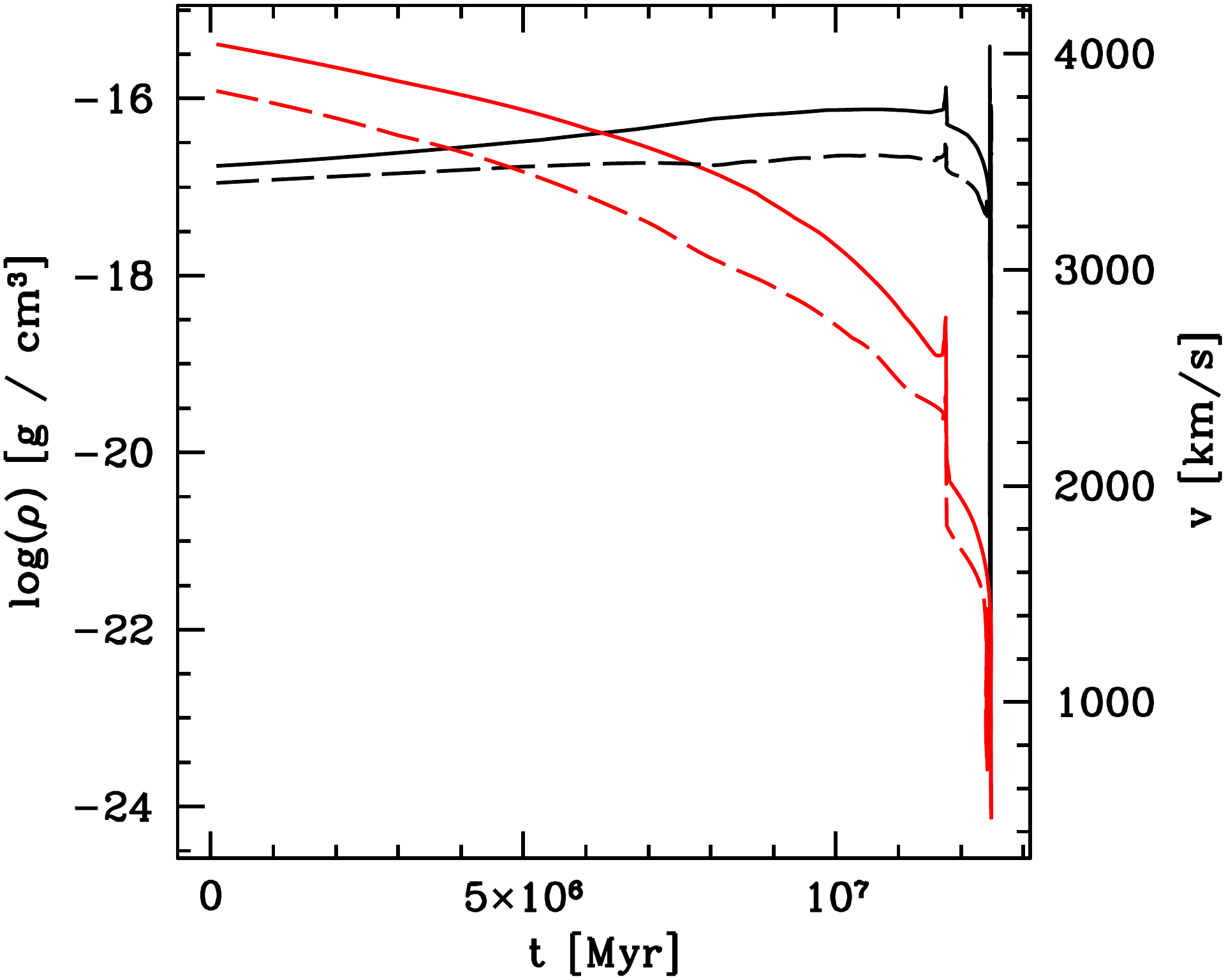}
\caption{\textit{Left panel:} Hertzsprung-Russell diagram of the $20\,\mathrm{M}_\odot$. Some key-points are indicated: the reaching of the critical velocity ($\Omega_\mathrm{crit}$), the end of the main sequence (End H-b), the departure from the critical velocity (end $\Omega_\mathrm{crit}$), the end of the central helium burning phase (end He-b) and the end of the central carbon burning phase (end C-b). \textit{Right panel:} Time evolution of the density (black lines) and velocity (red lines) of the stellar winds. The solid lines are for the polar regions and the dashed ones for the equator.}
\label{Fig20Msol}
\end{figure}

On Fig.~\ref{Fig20Msol}, we show on the left panel the Hertzsprung-Russell diagram for this model. Due to the low metallicity, the star remains on the blue side during its evolution. The critical velocity is reached already during the main sequence. When the star becomes redder, its radius inflates, and the surface velocity decreases. The star moves away from the critical velocity. The right panel shows the time evolution of the density (in red) and of the velocity (in black) of the stellar winds. The velocity of the wind is obtained with the relations of \citet{Kudritzki2000a}. The density is computed with the local mass flux \citep[with mass loss rate recipe of][]{Vink2000a,Vink2001a} and the velocity.The value for the poles (solid lines) are greater than the equatorial one (dashed lines), due to the von Zeipel theorem.

\section{Circum-stellar medium simulations}

The simulations of the CSM around the $20\,\mathrm{M}_\odot$ model discussed in the previous section are performed with the A-Maze hydrodynamical code \citep{Walder2000a}, which is a parallelised, adaptive mesh refinement code. The simulations start with a homogeneous medium, with the inner boarding conditions given by the wind data computed previously (they are thus time dependant). The chemical composition of the medium is also followed (for the main elements, H, He, C, N, and O). One set of simulations is performed in full 3D and cover the first few hundred thousands of years. The other set is done assuming an axial symmetry, and cover the whole lifetime of the star.

\begin{figure}
\begin{center}
\includegraphics[width=.9\textwidth]{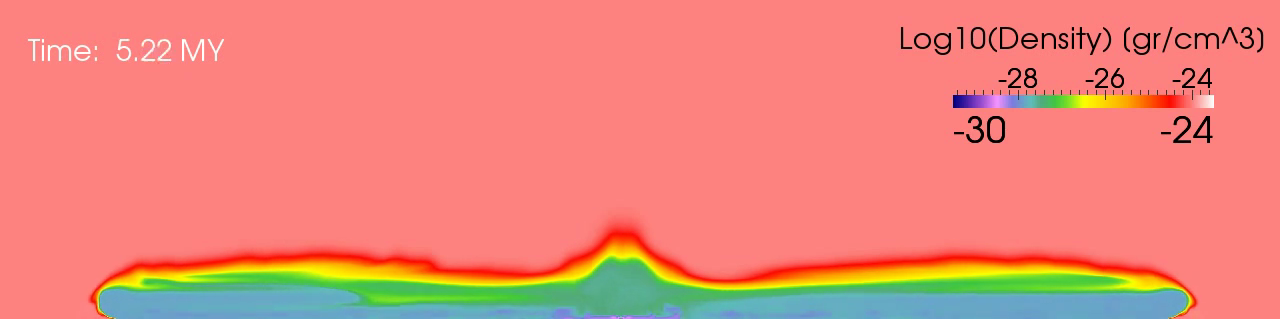}\\
\includegraphics[width=.9\textwidth]{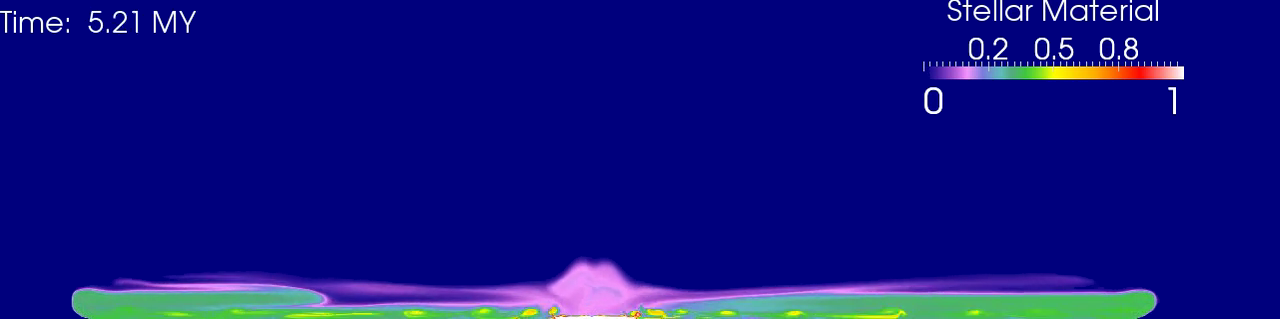}\\
\includegraphics[width=.9\textwidth]{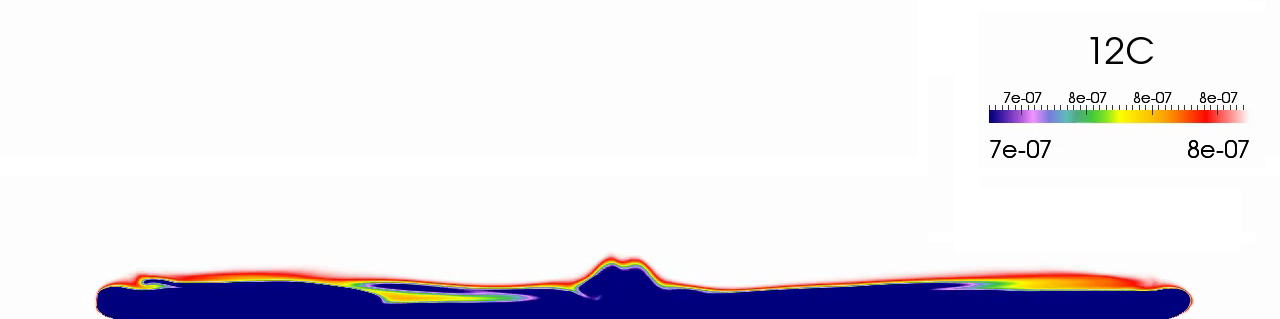}
\end{center}
\caption{\textit{Top panel:} Density around the star near the middle of the MS. The rotation axis is directed along the $x$-axis. The physical dimension along the $x$-axis is roughly $30\,\mathrm{pc}$. \textit{Medium panel:} Same as above panel, but for the fraction of material coming from the stellar winds. \textit{Bottom panel:} Abundance (mass fraction) of $^{12}$C surrounding the central star.}
\label{FigHydro2d}
\end{figure}

The main results are illustrated on Fig.~\ref{FigHydro2d}. Let us mention the following points:
\begin{itemize}
\item First of all, the stellar winds carve a strongly anisotropic cavity in the CSM, aligned with the rotation axis of the star. The density in the cavity is much smaller than in the ISM. Note also that the anisotropic shape of the cavity is not due to the 2D treatment, it also develops in the 3D simulations performed.
\item The cavity is progressively filled with the material of the star. After $\sim 5.2\,\mathrm{Myr}$, the fraction of matter coming from the star in the cavity is roughly $0.4$. The medium in the cavity is not homogeneous, and is turbulent.
\item Due to a strong mixing throughout the star, the stellar surface is progressively enriched by the products of hydrogen burning: the hydrogen abundance decreases, the helium one increases. Due to CNO-cycle, the carbon and oxygen abundances decrease while nitrogen abundance increases. These trends are clearly visible in the cavity dug by the winds.
\end{itemize}

At the end of the nuclear life of a fast rotating massive star, the CSM is thus probably strongly shaped by the stellar winds. The explosion of the central star will thus occur in a non-isotropic medium. The next step of this work will be to understand how such a medium can affect the explosion itself. This is particularly interesting in the frame of the long-soft Gamma Ray Bursts progenitors, which are thought to be fast rotating, low metallicity stars \citep[see e.g.][]{Yoon2006a}.

To improve our simulations, we are performing some additional computations: a) To check whether the aspect ratio of the cavity is affected by the Cartesian grid used in these preliminary simulations, the A-Maze code was modified, allowing for other (AMR) grid geometries, particularly spherical grids. b) Improving the equation of state, by adding the effects of partial ionisation of hydrogen and helium due to the strong ionising flux of the star. c) Starting from the 2D simulations, we also plan to perform 3D simulations of the most interesting phases of the evolution of the CSM, in order to check the 2D results and have a more detailed understanding of the CSM.

\bibliography{MyBiblio}

\end{document}